\documentclass{webofc}
\usepackage[varg]{txfonts}   
\usepackage{pythonhighlight}
\usepackage{hyperref}
\begin{document}
\title{The Scikit HEP Project - overview and prospects}
%
%

\author{\firstname{Eduardo} \lastname{Rodrigues}\inst{1}\fnsep\thanks{\email{eduardo.rodrigues@liverpool.ac.uk}} \and
        \firstname{Benjamin} \lastname{Krikler}\inst{2}\fnsep\thanks{\email{b.krikler@bristol.ac.uk}} \and
        \firstname{ Chris} \lastname{Burr}\inst{3}\fnsep\thanks{\email{christopher.burr@cern.ch}} \and
        \firstname{Dmitri} \lastname{Smirnov}\inst{4}\fnsep\thanks{\email{dsmirnov@bnl.gov}} \and
        \firstname{Hans} \lastname{Dembinski}\inst{5}\fnsep\thanks{\email{hans.dembinski@tu-dortmund.de}} \and
        \firstname{Henry} \lastname{Schreiner}\inst{6}\fnsep\thanks{\email{henry.fredrick.schreiner@cern.ch}} \and
        \firstname{Jaydeep} \lastname{Nandi}\inst{7}\fnsep\thanks{\email{nandi.jaydeep296@gmail.com}} \and
        \firstname{Jim} \lastname{Pivarski}\inst{8}\fnsep\thanks{\email{pivarski@princeton.edu}} \and
        \firstname{Matthew} \lastname{Feickert}\inst{9}\fnsep\thanks{\email{matthew.feickert@cern.ch}} \and
        \firstname{Matthieu} \lastname{Marinangeli}\inst{10}\fnsep\thanks{\email{matthieu.marinangeli@epfl.ch}} \and
        \firstname{Nick} \lastname{Smith}\inst{11}\fnsep\thanks{\email{nick.smith@cern.ch}} \and
        \firstname{Pratyush} \lastname{Das}\inst{12}\fnsep\thanks{\email{reikdas@gmail.com}}
}

\institute{University of Liverpool
\and
           University of Bristol
\and
           CERN
\and
           BNL
\and
           Technical University Dortmund
\and
           Princeton University
\and
           National Institute of Technology, Silchar
\and
           Princeton University
\and
           University of Illinois at Urbana Champaign
\and
           EPFL, Lausanne
\and
           FNAL
\and
           Institute of Engineering and Management, Kolkata
          }

\abstract{%
Scikit-HEP is a community-driven and community-oriented project with the goal of providing an ecosystem
for particle physics data analysis in Python.
Scikit-HEP is a toolset of approximately twenty packages and a few “affiliated” packages.
It expands the typical Python data analysis tools for particle physicists.
Each package focuses on a particular topic, and interacts with other packages in the toolset, where appropriate.
Most of the packages are easy to install in many environments;
much work has been done this year to provide binary “wheels” on PyPI and conda-forge packages.
%
%
The Scikit-HEP project has been gaining interest and momentum, by building a user and developer community
engaging collaboration across experiments.
Some of the packages are being used by other communities, including the astroparticle physics community.
An overview of the overall project and toolset will be presented, as well as a vision for development and sustainability.
}
\maketitle

\section{Introduction}
\label{intro}
\begin{figure}[h]
\centering
\includegraphics[width=8cm,clip]{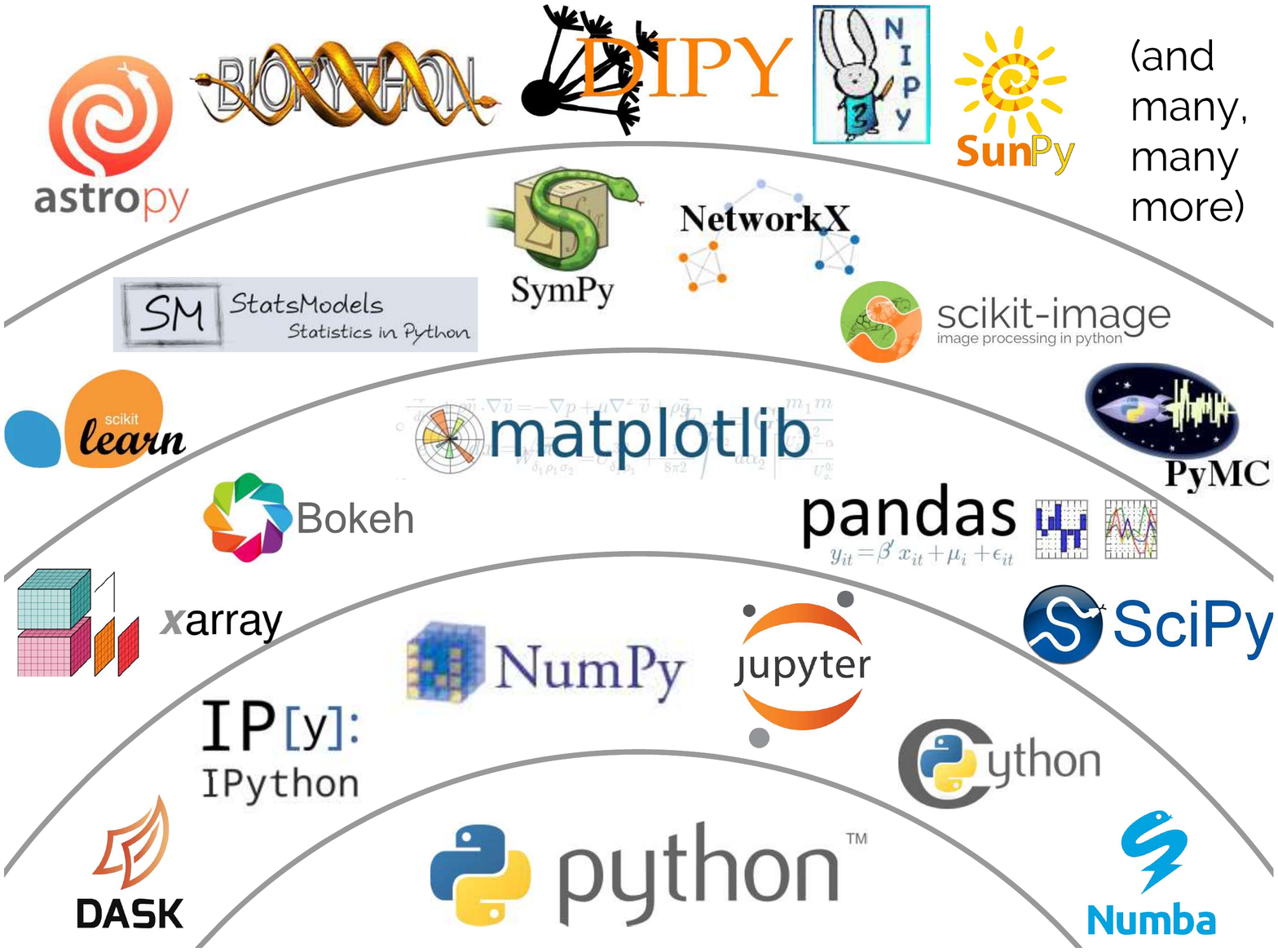}
\caption{Schematic view of the Python scientific software ecosystem.
Figure taken from Jake VanderPlas's presentation at the PyCon 2017 conference~\cite{JakeVanderPlas}.}
\label{fig-JakeVanderPlas}
\end{figure}

Python is an ever more popular programming language across a broad range of communities, notably in Data Science.
Outside High Energy Physics (HEP), the Python scientific ecosystem is built atop the "building blocks"
of the SciPy ecosystem of open-source software for mathematics, science, and engineering~\cite{SciPy}.
Figure~\ref{fig-JakeVanderPlas} provides a good visual illustration of the ecosystem,
which grows from foundational libraries all the way to domain-specific projects such as Astropy~\cite{astropy}.
The ecosystem provides tools for data manipulation, visualisation, statistics, machine learning, etc.

Traditionally, HEP has been evolving in a rather disjoint ecosystem based on the C++ ROOT data analysis framework~\cite{ROOT}.
Same as for the Python scientific ecosystem, it provides tools for data manipulation and modeling, for fitting,
for statistics and machine learning applications. But it is a \textit{toolkit} rather than a \textit{toolset}, with an interface that is not
that natural for Python --- via bindings it provides.

Various initiatives tried to link both HEP and non-HEP worlds, at least for specific tasks.
Unfortunately, the libraries were very largely developed by a single author and sustainability became quickly an issue, especially that many such authors left the field. Also, community adoption did not stick.
We believed that a more generalised effort, domain-specific oriented, was the way forward, and this gave rise to the Scikit-HEP project in late 2016.
It is only in 2018 that community adoption started to take up, with several project packages attracting much attention.
We are proud to mention that several collider (Belle II, CMS) and non-collider (KM3NeT) experiments officially use some of Scikit-HEP in their external dependencies,
as do other software projects (Coffea, zfit).

The project had been presented at the CHEP 2018 conference~\cite{CHEP2018}. The present report superseeds Ref.~\cite{CHEP2018}
and presents the status of the project, which evolved considerably since CHEP 2018.

\section{Scikit-HEP project overview}
\label{project}
The Scikit-HEP project~\cite{Scikit-HEP-project} is a community-driven and community-oriented effort with the aim
of providing Particle Physics at large with a \textit{toolset} ecosystem for data analysis in Python.
It does not attempt in any way to provide a replacement for the Python ecosystem based on the SciPy suite;
it rather builds on its foundational libraries providing core and common tools for the HEP community.
The grand plan of the project can be summarised in the following points:
\begin{itemize}
\item Create an ecosystem for particle physics data analysis in Python.
\item Improve the interoperability between HEP tools and the scientific ecosystem in Python.
\item Expand the typical toolset for particle physicists with high-standards, well-documented and easily installable domain-specific packages.
\item Build a community of developers and users, having sustainability in mind.
\item Improve discoverability of (domain specific) relevant tools.
\end{itemize}

\begin{figure}[h]
\centering
\includegraphics[width=\textwidth,clip]{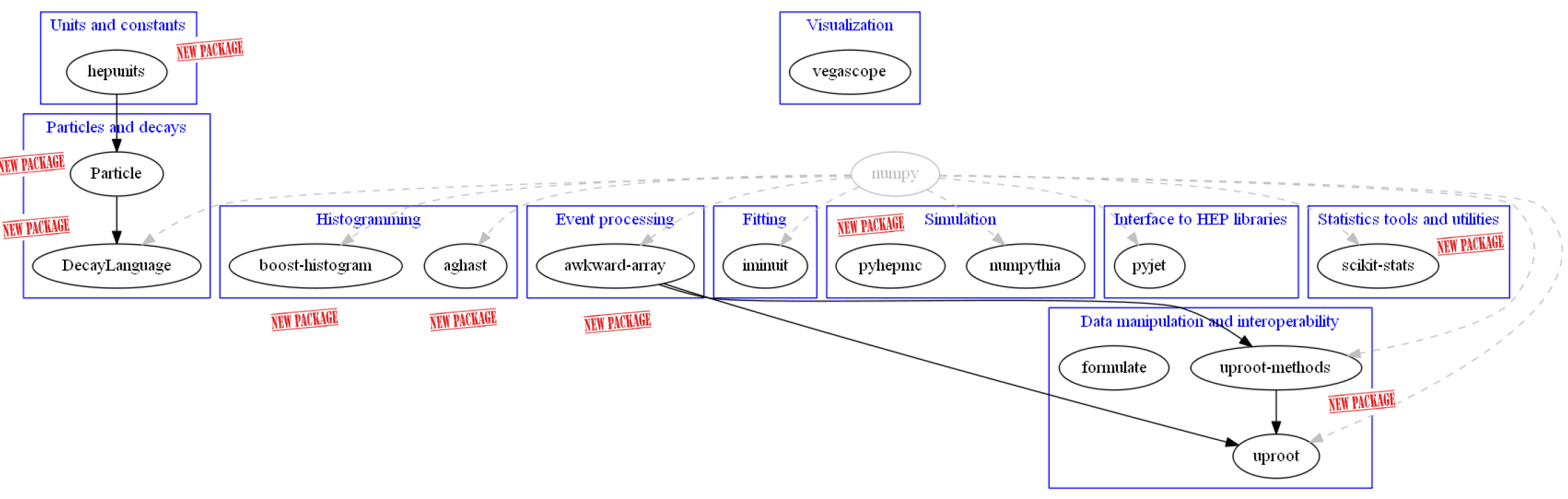}
\caption{Overview of (most of) the packages making the Scikit-HEP toolset.
For a sense of evolution all packages whose first release came out after the CHEP 2018 conference are marked as "new package".
All GitHub repositories can be found at the location \url{https://github.com/scikit-hep}.}
\label{fig-Scikit-HEP}
\end{figure}

The Scikit-HEP toolset is depicted (to a large extent) in figure~\ref{fig-Scikit-HEP}.
Some of the packages found in the GitHub organisation, such as the well-known packages \pyth{root_numpy}~\cite{root-numpy}
and \pyth{root_pandas}~\cite{root-pandas}, pre-dating the project, are not described in this report.
They are nevertheless part of the project, but largely deprecated by the new and more versatile packages
\pyth{uproot}~\cite{uproot} and \texttt{awkward-array}~\cite{awkward-array}, see below.
More importantly, it should be emphasised that most of the packages presently constituting the Scikit-HEP toolset are relatively new,
having been released for the first time after the CHEP 2018 conference; these are marked as "new package" in figure~\ref{fig-Scikit-HEP}.

The remainder of this report provides a whirlwind tour of the main packages.

\vspace*{0.4cm}

\section{Whirlwind tour of Scikit-HEP packages}
\label{packages}
The obvious "point of entry" to the HEP ecosystem is via ROOT files. These can be natively and trivially read with the pure Python I/O \pyth{uproot} package~\cite{uproot},
whose only dependencies are NumPy and Python libraries to deal with compression and decompression.
The package is installable via \pyth{pip} or \pyth{conda} on virtually any computer, straighforwardly since it does \textit{not} depend on ROOT. It has been a runaway success with over 15000 downloads per month.

\pyth{Uproot} bridges the ROOT and the NumPy-based ecosystems in its most simple incantation. Indeed a ROOT \pyth{TTree} object is read as a dictionary of arrays as follows:

\begin{python}
>>> import uproot
>>> t = uproot.open("Zmumu.root")["events"]
>>> t.arrays(["px1","px2","py1","py2","M"])
{b'px1': array([-41.19528764,  35.11804977,  35.11804977, ...,  32.37749196,
         32.37749196,  32.48539387]),
 b'px2': array([ 34.14443725, -41.19528764, -40.88332344, ..., -68.04191497,
        -68.79413604, -68.79413604]),
 b'py1': array([ 17.4332439 , -16.57036233, -16.57036233, ...,   1.19940578,
          1.19940578,   1.2013503 ]),
 b'py2': array([-16.11952457,  17.4332439 ,  17.29929704, ..., -26.10584737,
        -26.39840043, -26.39840043]),
 b'M': array([82.46269156, 83.62620401, 83.30846467, ..., 95.96547966,
        96.49594381, 96.65672765])}
\end{python}

\noindent
A plethora of other manipulations are available, such as iteration over several files and opening a remote file.
Note that \pyth{uproot} is able since a few months to write simple ROOT objects such as \pyth{TTree}.

HEP analyses typically involve the manipulation of complex data structures that can be
\begin{itemize}
\item Variable length lists (jagged/ragged);
\item Deeply nested (record structure);
\item And of different data types in the same list (heterogeneous).
\end{itemize}
The \texttt{awkward-array}~\cite{awkward-array} package provides a way to analyse these variable-length tree-like data in Python
by extending Numpy's idioms from flat arrays to arrays of data structures.
The package is being reimplemented in C++, with a simpler interface and less limitations, based on acquired experience and user feedback;
the developments are taking place at \url{https://github.com/scikit-hep/awkward-1.0}.

The analysis of datasets (processed \textit{e.g.} with the two packages just described) typically involves data aggregations;
these are most often in the form of one-dimensional histograms.
Indeed, histogramming is central in any analysis workflow and has received much attention.
The package \texttt{boost-histogram}~\cite{boost-histogram} bundles the Python bindings for the performant C++14 multi-dimensional templated header-only library Boost.Histogram~\cite{Boost.Histogram}, albeit with a Pythonic API.
Histograms can be defined in a very versatile way owing to the extensive types of axes (regular, variable, circular axes, etc.)
and storages (interger, double, weighted values) defined, in multi dimensions.
The package provides methods for selecting, rebinning, and projecting into lower-dimensional space.
It is a high-performance histogramming package naturally talking to the NumPy ecosystem. Its interface is simple and user-friendly:

\begin{python}
>>> import boost_histogram as bh
>>>
>>> import numpy as np
>>> import matplotlib.pyplot as plt
>>>
>>> hist = bh.Histogram(bh.axis.Regular(30, 0, 2*np.pi, circular=True))
>>> hist.fill(np.random.uniform(0, np.pi*4, size=300))
>>>
>>> plothist = lambda h: plt.bar(h.axes[0].centers, h, width=h.axes[0].widths)
>>> ax = plt.subplot(111, polar=True)
>>> plothist(hist);
\end{python}

\noindent
This code snippet produces this output (figure~\ref{fig-ex-boost-histogram}):

\begin{figure}[h]
\centering
\includegraphics[width=0.5\textwidth,clip]{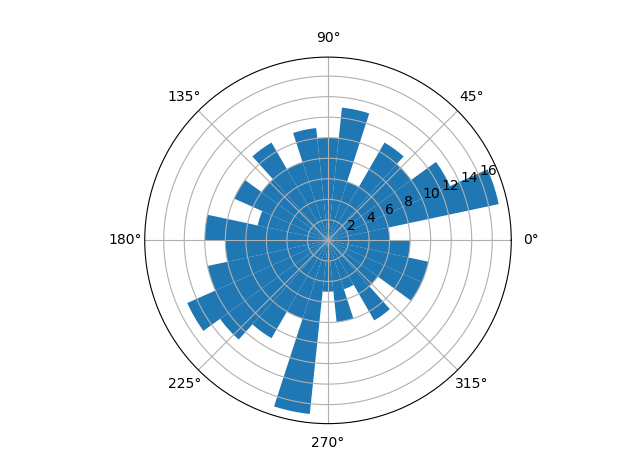}
\caption{Example of a \texttt{boost-histogram} histogram with circular axes.}
\label{fig-ex-boost-histogram}
\end{figure}

Continuing a common flow of work in a HEP analysis, it is important to mention the \pyth{iminuit} fitting package~\cite{iminuit},
the Python interface (bindings) to the \texttt{Minuit2} C++ package~\cite{Minuit2} used across particle physics.
\texttt{Minuit2} is most commonly used for likelihood fits of models to data, and to get model parameter error estimates from likelihood profile analysis.
The bindings constitute an important building block for sophisticated data modelling packages. It is used in other HEP packages and in various astroparticle physics packages.

The Scikit-HEP project contains other utility packages such as:

\begin{itemize}
\item The \pyth{Particle} package~\cite{Particle}: a Pythonic interface to the Particle Data Group (PDG) particle data table and Monte Carlo particle identification codes,
with a multitude of goodies such as powerful and flexible searches as one-liners.
\item The \pyth{DecayLanguage} package~\cite{DecayLanguage}: tools to parse decay files and programmatically manipulate them, query and display information;
classes for a universal representation of particle decay chains.
\item The \pyth{numpythia}~\cite{numpythia} and \pyth{pyjet}~\cite{pyjet} packages:
they provide interfaces between NumPy and the popular Pythia~\cite{Pythia} event generator and the FastJet~\cite{FastJet} jet finding algorithm, respectively.
\end{itemize}

\section{Outlook}
\label{outlook}
The Scikit-HEP project has been gaining much interest and momentum in the last couple of years.
Together with other projects, it is providing a modern and alternative ecosystem for HEP analysis, in Python.
The project is community-driven and community-oriented.
It is building a user and developer community engaging collaboration across experiments. This is crucial to ensure continuity and sustainability,
with a culture where the users of today are meant to become the developers of tomorrow.
Some of the project packages are being used by other communities, including the astroparticle physics community.

%
%
%

\end{document}